\documentclass[aps,pra,reprint,superscriptaddress]{revtex4-2}

\usepackage[english]{babel}
\usepackage{graphics}
\usepackage{todonotes}
\usepackage{graphicx}
\usepackage{amsfonts,amssymb,amsmath}
\usepackage{color}

\definecolor{darkblue}{rgb}{0.0,0.0,0.4}
\definecolor{darkgreen}{rgb}{0.0,0.4,0.0}
\definecolor{darkred}{rgb}{0.6,0.0,0.0}
\usepackage[bookmarksopen]{hyperref}
\hypersetup{colorlinks,linkcolor=darkblue,citecolor=darkblue,urlcolor=darkblue}

\begin{document}

\title{Self-bound dipolar droplets and supersolids\\in molecular Bose-Einstein condensates}
\author{Matthias Schmidt} 
\affiliation{5. Physikalisches  Institut  and  Center  for  Integrated  Quantum  Science  and  Technology,Universit\"at  Stuttgart,  Pfaffenwaldring  57,  70569  Stuttgart,  Germany}

\author{Lucas Lassabli\`ere}
\affiliation{Universit\'e Paris-Saclay, CNRS, Laboratoire Aim\'e Cotton, 91405 Orsay, France}

\author{Goulven Qu\'em\'ener}
\affiliation{Universit\'e Paris-Saclay, CNRS, Laboratoire Aim\'e Cotton, 91405 Orsay, France}

\author{Tim Langen}
\email{t.langen@physik.uni-stuttgart.de}

\affiliation{5. Physikalisches  Institut  and  Center  for  Integrated  Quantum  Science  and  Technology,Universit\"at  Stuttgart,  Pfaffenwaldring  57,  70569  Stuttgart,  Germany}

\begin{abstract}
We numerically study the many-body physics of molecular Bose-Einstein condensates with strong dipole-dipole interactions. We observe the formation of self-bound droplets, and explore phase diagrams that feature a variety of exotic supersolid states. In all of these cases, the large and tunable molecular dipole moments enable the study of unexplored regimes and phenomena, including liquid-like density saturation and universal stability scaling laws for droplets, as well as pattern formation and the limits of droplet supersolidity. We discuss a realistic experimental approach to realize both the required collisional stability of the molecular gases and the independent tunability of their contact and dipolar interaction strengths. Our work provides both a blueprint and a benchmark for near-future experiments with bulk molecular Bose-Einstein condensates. 
\end{abstract}

\maketitle

\section{Introduction}
Quantum fluctuations can stabilize dipolar Bose-Einstein condensates against their mean-field collapse~\cite{Bottcher2020,Petrov2015,Kadau2016,Wachtler2016}. This counterintuitive behavior leads to a rich phase diagram that contains, for example, self-bound quantum droplets~\cite{Baillie2016,Schmitt2016} and supersolid states~\cite{Bottcher2019,Roccuzzo2019,Tanzi2019,Chomaz2019}. While many aspects like
rotonic excitation spectra~\cite{Santos2003,Wilson2008,Bismut2012,Bisset2013,Chomaz2018,Hertkorn2019,Petter2019,Schmidt2021,Hertkorn2021SSD2D}, anisotropic superfluidity~\cite{Wenzel2018,Ferrier2018}, droplet formation~\cite{Bottcher2020,Schmitt2016,Chomaz2016}, crystallization in 1D~\cite{Guo2019,Natale2019,Tanzi2019a,Hertkorn2021,Tanzi2021,Ilzhoefer2021}, 2D~\cite{Kadau2016,Baillie2018,Schmidt2021,Norcia2021,Hertkorn2021SSD2D} and into more exotic patterns~\cite{Zhang2021phases,Hertkorn2021pattern}, have been extensively discussed for weakly-dipolar magnetic atoms, systematic studies for molecules, with their much larger and tunable electric dipole moments, have so far remained scarce. Here, we show that Bose-Einstein condensates of ground-state molecules are ideal candidates to further explore the rich phase diagrams of dipolar Bose gases in experiments~\cite{Bohn2017}.  

Our work is motivated by the recent extraordinary progress in the preparation of molecular ensembles at ultracold temperatures. This includes progress in magneto-association from ultracold atoms~\cite{Koehler2006}, which has recently lead to the creation of a collisionally stable degenerate Fermi gas~\cite{DeMarco2019,Valtolina2020,Matsuda2020,Duda2021}. It also includes progress in direct laser cooling of molecules, where the achievable phase space densities have increased by almost ten orders of magnitude over the last few years~\cite{Barry2014,Norrgard2016,Prehn2016,Truppe2017,Anderegg2018,Ding2020}. Further cooling for various molecular species is expected to be possible through collisional cooling~\cite{Reens2017,Segev2019,Son2020,Jurgilas2021,Li2021}. With this progress, a Bose-Einstein condensate (BEC) of strongly dipolar ground state molecules is now within experimental reach. 

Previous theoretical studies have investigated molecular Bose gases with dipolar interactions in various scenarios from the weakly to the strongly interacting
limit in lattices and in bulk systems~\cite{Lu2015,Buechler2007,Trefzger2009,Danshita2009,Pollet2010,Capogrosso2010, Baranov2012,Macia2016,Cinti2017,Kora2019}. Purely dipolar systems have also been used to study crystallisation~\cite{Astrakharchik2007,Mora2007,Knap2012}, localization~\cite{Yao2014} and topological states~\cite{Micheli2006}.

In this work we apply the extensive toolkit pioneered for the understanding of magnetic BECs to explore the many-body physics that can be observed with a future molecular BEC. Moreover, we discuss in detail the collisional stability of the molecules and show how the required interaction parameters for contact and dipolar interactions can realistically be achieved in experiments. 

In the following, we will focus on NaRb molecules, which have a permanent electric dipole moment of $d_p=3.2\,\mathrm{D}$~\cite{Guo2016,Ye2018,Guo2018prx}. This value corresponds to a dipolar interaction strength that is up to three orders of magnitude larger than the one in magnetic atoms like dysprosium or erbium. We emphasize that, given the universal nature of the collisional processes on the one hand~\cite{Gonzales2017,Lassabliere2018} and the the scaling properties of the extended Gross-Pitaevskii equation on the other hand~\cite{Hertkorn2021pattern}, the results presented in the following can easily be generalized to other molecules, including typical laser-coolable species.

\section{Extended Gross-Pitaevskii theory}
We model the molecular BEC using the extended Gross-Pitaevskii-Equation (eGPE)~\cite{Wachtler2016,Bottcher2020,Lahaye2009}
\begin{equation}
	i\hbar\, \partial_t \psi = \left( \hat{H}_0 + g\left|\psi\right|^2 + \Phi_\mathrm{dd} +g_\mathrm{qf}\left|\psi\right|^3\right)\psi,
\label{eq:eGPE}
\end{equation}
where the wavefunction $\psi\equiv\psi(\boldsymbol{r},t)$ is normalized to the molecule number $N= \int \left| \psi\right|^2 \mathrm{d}^3r$ and  $\hat{H}_0=  - \frac{\hbar^2 \Delta}{2m} +V_\mathrm{ext}(\boldsymbol{r})$ describes the motion of a single molecule in an external harmonic trapping potential $V_\mathrm{ext}(\boldsymbol{r})=\frac{m}{2}(\omega^2_x x^2+\omega^2_y y^2 + \omega^2_z z^2)$. Here, $m$ denotes the molecular mass of the molecules and $\omega_{x,y,z}$ are the trapping frequencies in the spatial directions $\boldsymbol{r}=(x,y,z)$. 

The short-range contact interaction between the molecules is characterized by the coupling constant $g= 4\pi\hbar^2 a_\mathrm{s}/m$, where $a_\mathrm{s}$ is the s-wave scattering length. As molecules can be lost from two-body processes, $a_\mathrm{s}$ is in general a complex quantity, with the imaginary part related to the losses. We consider here the case with no losses so that the imaginary part of $a_\mathrm{s}$ is set to zero. We will see in Sec. V how to do so. 

The mean-field potential $\Phi_\mathrm{dd}$ of the dipole-dipole interaction (DDI) is given by $\Phi_\mathrm{dd}= g_\mathrm{dd}\int \mathrm{d}^3r^\prime |\psi(\boldsymbol{r}^\prime)|^2 U(\boldsymbol{r}-\boldsymbol{r}^\prime)$ with $g_\mathrm{dd}=4\pi\hbar^2 a_\mathrm{dd}/m$ being the DDI coupling constant~\cite{Lahaye2009}. Here, we have introduced the dipolar length $a_\mathrm{dd}=d^2m/12\pi\hbar^2\epsilon_0$ to characterize the strength of the dipolar interaction, with $\epsilon_0$ the vacuum dielectric constant. For molecules, the dipolar length is mediated by the induced dipole moment $d$, which depends on the value of an applied static electric field $E_{dc}$ that polarizes the molecules. This induced dipole moment represents the expectation value of the permanent dipole moment $d_p$ in a given rotational state along the electric field axis, taken as the quantization axis in the $z$-direction. In this configuration, the geometric part of the interaction potential is given by $U(\boldsymbol{r})=(3/4\pi)(1-3z^2/r^2)/r^3$, with $r=|\boldsymbol{r}|$.

Typical values for $a_\mathrm{dd}$ realized in magnetic atoms are $16\,a_0$ for chromium~\cite{Lahaye2009}, $65\,a_0$ for erbium~\cite{Chomaz2016} and $130\,a_0$ for dysprosium~\cite{Bottcher2019droplet}, with the Bohr radius $a_0$. For NaRb molecules, $a_\mathrm{dd}$ can, in principle, be tuned to any value between zero (unpolarized molecules) and $1.1\,\times\,10^5\,a_0$ (fully polarized molecules), depending on the value of the applied external electric field.  

We further consider beyond mean-field contributions from quantum fluctuations, which play a crucial role in dipolar systems~\cite{Lima2011,Petrov2015}. We include them in the mean-field description using a local density approximation. The corresponding Lee-Huang-Yang (LHY) correction is given by $g_\mathrm{qf}\left|\psi\right|^3$, with $g_\mathrm{qf} = (32/3\sqrt{\pi})ga_\mathrm{s}^{3/2}(1+3\varepsilon^2_\mathrm{dd}/2)$, and $\varepsilon_\mathrm{dd}= a_\mathrm{dd}/a_\mathrm{s}$ denoting the relative dipolar strength. 

We find the ground states of this system, both in free space and in the presence of an external trapping potential, using imaginary time evolution. Details of our numerical procedure to solve Eq.~\ref{eq:eGPE} have been discussed in detail in previous works~\cite{Hertkorn2019,Hertkorn2021SSD2D,Mennemann2015}. 

The richness of this system is a result of the different scaling of the individual terms in Eq.~\ref{eq:eGPE} with density. This leads to a complex interplay of mean-field contact and dipolar interactions, and their respective beyond mean-field LHY corrections. In particular, fine-tuned situations can arise where the small repulsive beyond-mean-field contributions dominate over large attractive mean-field terms, leading to a stabilization of systems that would collapse on the pure mean-field level. Naturally, the resulting stability and phase diagrams depend crucially on the exact magnitude of the beyond mean-field LHY contributions. The validity of approximations made to incorporate these contributions is a topic of ongoing investigations~\cite{Bottcher2020}. Our study is thus not only motivated by the search for novel states of matter inaccessible to existing experiments, but also by the need to identify situations where the validity of the eGPE can be benchmarked systematically. 

\begin{figure}[tb]
\centering
\includegraphics[width=0.48\textwidth]{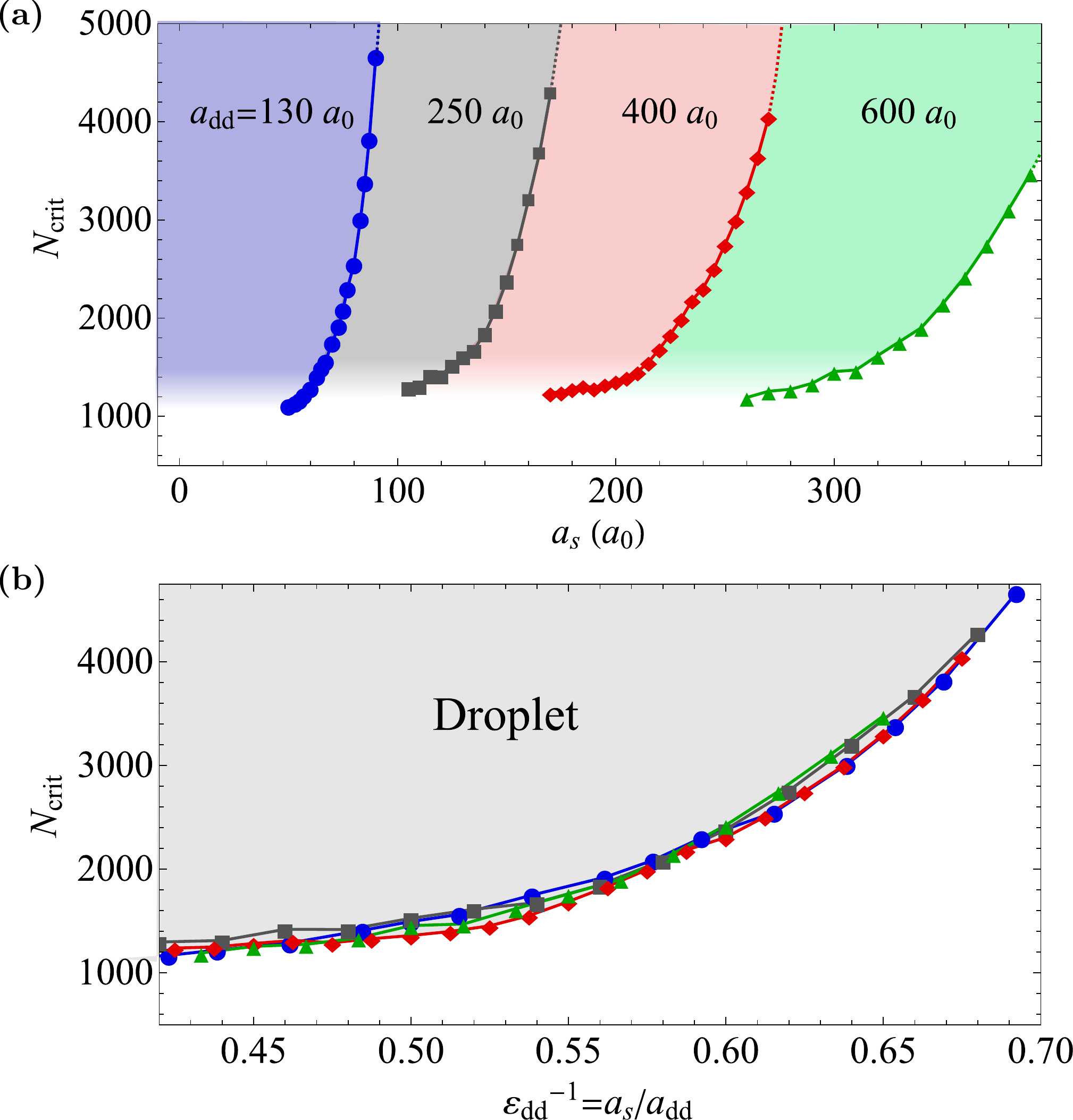}
\caption{Stability of self-bound molecular droplets. (a) Critical molecule number $N_\mathrm{crit}$ for several values of the dipolar length. For a fixed dipolar length the critical molecule number increases as a function of the scattering length $a_s$. The first value, $a_\mathrm{dd}=130\,a_0$, coincides with the value in dysprosium, the most magnetic atomic species. For larger $a_\mathrm{dd}$, the required scattering length $a_s$ for stabilization increases. 
(b) Rescaling shows that $N_\mathrm{crit}$ is a universal function of the dipolar strength $\varepsilon_\mathrm{dd}$~\cite{Baillie2016}. As $N_\mathrm{crit}$ scales strongly with $\varepsilon_\mathrm{dd}$, measurements of this universal function can be a sensitive test of the microscopic collisional processes and the underlying many-body physics.}
\label{fig:criticalatomnumber}
\end{figure}

\section{Self-bound droplets}
We start by investigating a molecular BEC in free space without the presence of an external trapping potential. In this configuration a regular BEC is an unstable solution of the eGPE, since the overall mean-field energy is positive. However, for suitable interaction strengths and molecule numbers above a certain critical number $N_\mathrm{crit}$ a self-bound droplet can emerge, which is stabilized by the repulsive LHY contributions~\cite{Bottcher2020,Petrov2015,Baillie2016,Schmitt2016,Chomaz2016}.\\

In a first step we investigate the influence of the dipolar length $a_\mathrm{dd}$ on $N_\mathrm{crit}$. To find $N_\mathrm{crit}$ numerically we follow the established procedure to first compute the droplet ground state at a molecule number of $N= 15 \times 10^3$, which is much larger than $N_\mathrm{crit}$ for all parameters considered~\cite{Bottcher2019droplet}. In a second step, we temporarily include three-body losses by adding the term $\hat{H}_{L_{3}}= -i\hbar L_3 \left|\psi\right|^4/2$ to the Hamiltonian of Eq.~\ref{eq:eGPE}, with $L_3$ a 3-body loss coefficient. We then simulate the time evolution of the system under the influence of these losses to identify the molecule number where the droplet is no longer stable. 

Apart from this numerical procedure, and in analogy with the absence of two-body losses, we will assume that three-body losses are zero. This assumption is supported by recent experiments with KRb molecules~\cite{Matsuda2020,Li2021} and will be further discussed in Sec. VI.  

In Fig.~\ref{fig:criticalatomnumber}a, we present the results for several moderately large values of $a_\mathrm{dd}$ up to $600\,a_0$. While these values significantly exceed the values realized in the most magnetic atoms, they are easily achievable in many molecules. 

\begin{figure}[tb]
\centering
\hspace{-7pt}
\includegraphics[width=0.48\textwidth]{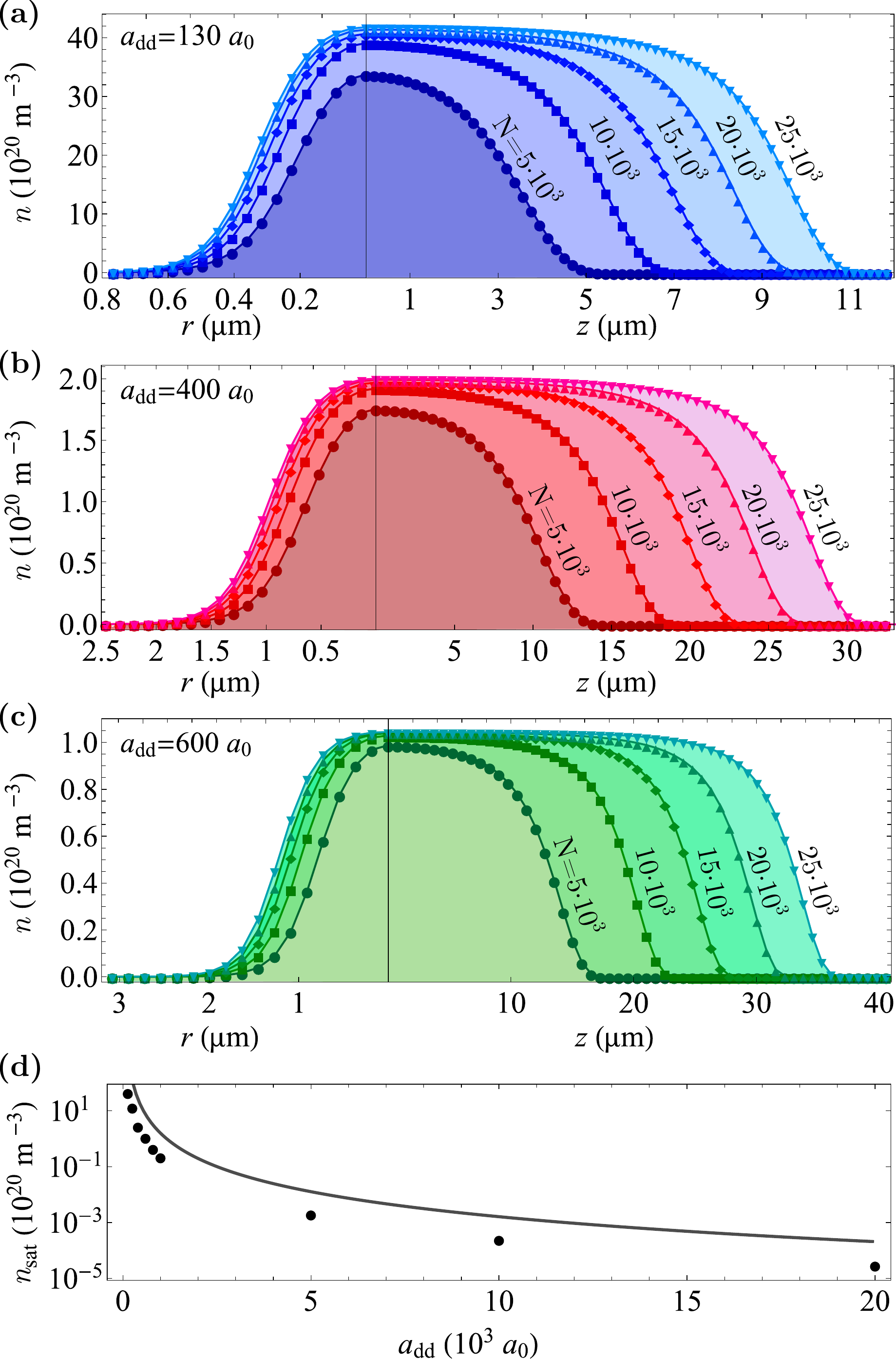}

\caption{Droplet density saturation. (a,b,c) Density distributions $n(r,z)$ of self-bound droplets for several molecule numbers $N$ and dipolar lengths $a_\mathrm{dd}$, plotted along the droplets' radial ($r$) and axial ($z$) directions. The droplets are strongly elongated, with the axial size being an order of magnitude larger than the radial size. In these plots $\varepsilon_\mathrm{dd}=2$, and the axial direction is along the polarization direction of the molecules. The density in the center of the droplet increases for larger $N$, until a saturation density $n_\mathrm{sat}$ is reached and the incompressibility of the droplet prevents a further increase of the density. This leads to a characteristic flat-top density distribution. For increasing $a_\mathrm{dd}$ (b,c) the saturation density is reached already for smaller molecule numbers and the density distribution develops larger saturated regions in the center. (d) We study the decrease of the saturation density $n_\mathrm{sat}$ with increasing dipolar strength up to large values exceeding $a_\mathrm{dd} \sim 20\times 10^3\,a_0$. Notably, this decrease can qualitatively also be understood by a variational model (solid line)~\cite{Ferrier-Barbut2016}.}
\label{fig:dropletdensities}
\end{figure}

For a specific dipolar length $a_\mathrm{dd}$ the critical molecule number increases as a function of scattering length, and higher values of $a_\mathrm{dd}$ require higher scattering lengths for stabilization. This can intuitively be understood as larger values of $a_\mathrm{dd}$ lead to larger dipolar attraction within the elongated droplets. In order to achieve stability, this attraction needs to be compensated by a higher repulsion from the contact interaction. As an example, for $N=2000$ and the values of $a_\mathrm{dd}$ considered, values of the scattering length on the order of $a_s\sim\,0.6\,a_\mathrm{dd}$ are required for stability. 

Our results highlight how independent tuning of the dipolar strength and the s-wave scattering length will be imperative for the realization of molecular droplets. However, while the critical molecule number changes with both $a_\mathrm{dd}$ and $a_s$, it is expected to be a universal function of their ratio $\varepsilon_\mathrm{dd}$ within eGPE theory. This behavior is known since the early work on quantum droplets~\cite{Baillie2016} and is reproduced well by our simulations, as shown in Fig.~\ref{fig:criticalatomnumber}b. Molecular BECs, despite being much more strongly dipolar than magnetic BECs, will thus allow the exploration of a similar region of this universal curve. This provides an interesting setting to cross-validate results. In particular, previous experiments with magnetic dysprosium BECs have shown indications of a systematic shift from the universal prediction~\cite{Bottcher2019droplet}. This shift has been attributed, for example, to finite temperature corrections to the scattering problem~\cite{Oldziejewski2016} and to approximations in the derivation of the LHY term in the eGPE~\cite{Bottcher2020}. Performing similar experiments with molecules over a large range of different $a_\mathrm{dd}$ will thus be a powerful benchmark both for the underlying collisional theories and the LHY term in the eGPE. 

\begin{figure*}[tb]
\centering
\includegraphics[width=0.98\textwidth]{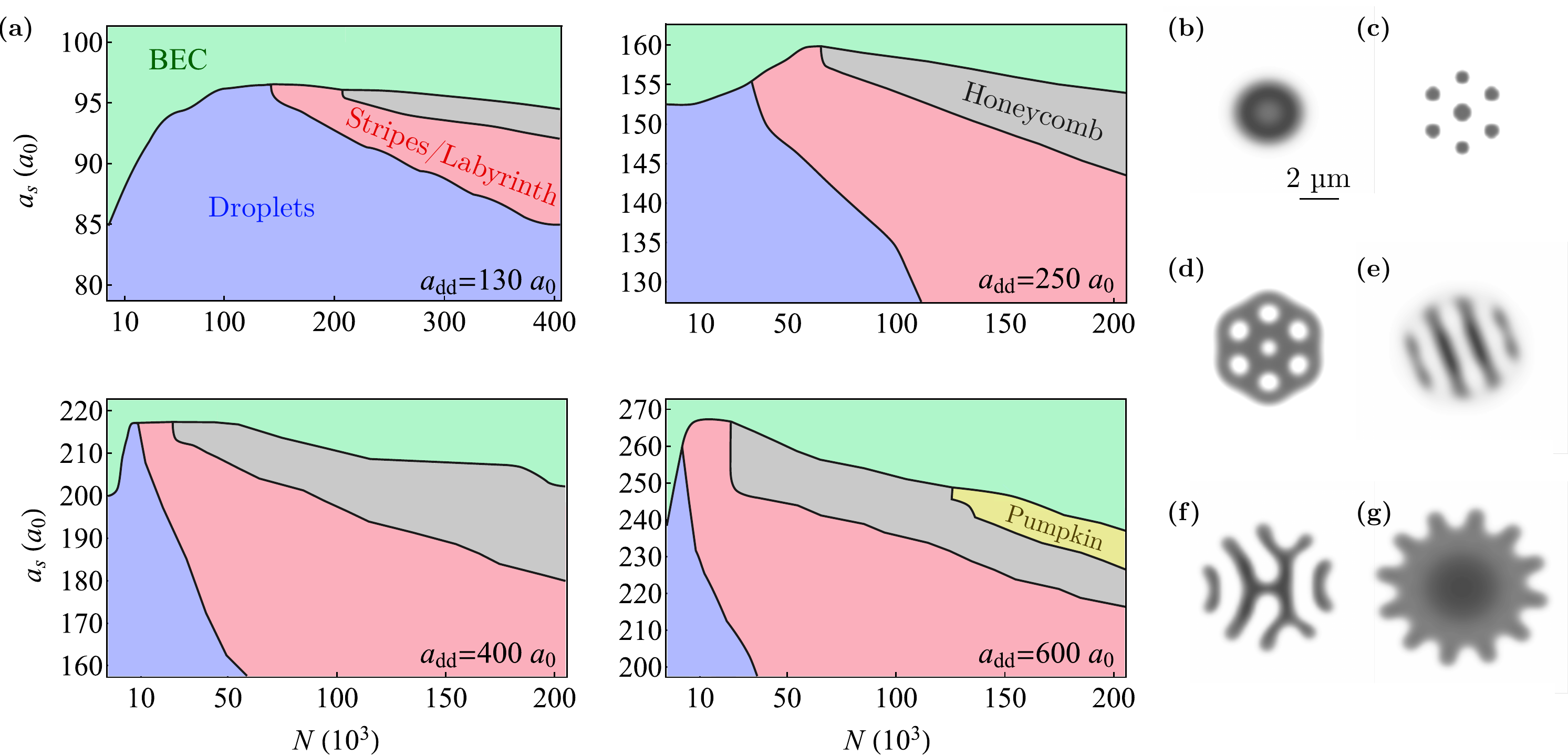}  
\caption{Supersolid states of trapped molecular BECs. (a) Phase diagrams for various dipolar lengths $a_\mathrm{dd}$. For each $a_\mathrm{dd}$, bloodcell-shaped BECs (b) can be transformed into various symmetry-broken states including supersolid and isolated droplet arrays (c), honeycomb (d), stripe and labyrinth (e,f), and pumpkin patterns (g), which each exhibit different sizes and shapes depending on the parameters. Note how the example honeycomb pattern in (d) exhibits a density distribution that is the exact inverse of the droplet state (c). The parameters required to realize the individual patterns shift towards smaller molecule numbers $N$ with increasing dipolar length $a_\mathrm{dd}$, bringing the corresponding states within realistic experimental reach.  
}
\label{fig:phasediagram}
\end{figure*}

Next, we investigate the density distribution of the droplets. A striking feature of quantum droplets is the liquid-like saturation of their central density, in analogy to the behavior of water or helium droplets~\cite{Toennies2001,Volovik2003,Dalfovo2001}, but at orders of magnitude lower density. In magnetic quantum gases, reaching this saturation limit requires atom numbers significantly beyond current experimental capabilities. As a consequence, density saturation has so far not been observed in equilibrium.

In Fig.~\ref{fig:dropletdensities}a we show the density distribution of droplets for various dipolar lengths and molecule numbers. The droplet size increases strongly with increasing dipolar length $a_\mathrm{dd}$ and molecule number $N$. The increase along the polarization axes of the dipoles is larger than for the radial size of the droplets. Such large droplets constitute an ideal platform to study questions of elementary excitations~\cite{Petrov2015,Pal2021}, self-bound vortex droplets~\cite{Lee2020}, and thermalization~\cite{Wenzel2018b}. These questions are well studied for helium droplets~\cite{Toennies2001,Volovik2003}, but remain experimentally unexplored for dipolar droplets. 

The characteristic saturation of the density is clearly visible as a plateau that emerges in the density distribution in the center of the droplet. In particular, for increasing $a_\mathrm{dd}$ one observes that the saturation density is reached already at much lower molecule numbers. We have confirmed that this behavior continues up to much stronger dipolar interactions exceeding $a_\mathrm{dd} = 2\,\times\,10^4\,a_0$. As shown in Fig.~\ref{fig:dropletdensities}d, this can also qualitatively be explained using a variational ansatz~\cite{Ferrier-Barbut2016}. Notably, from further simulations including also a weak trapping potential, we find that once in the saturated regime, the peak density only depends on $a_s$ and $a_\mathrm{dd}$, but not significantly on the trapping geometry. 

The limitation of the density to values that are significantly below the ones typically encountered in magnetic atoms has several  practical consequences. First, it guarantees that the quantum depletion remains small, $na_s^3\ll 1$ and $na_\mathrm{dd}^3\ll 1$, throughout the parameters studied here. While the gases studied are strongly dipolar, they thus remain in a regime, where the description with the eGPE is expected to be valid. It will be interesting to explore experimentally for how large values of $a_\mathrm{dd}$ this is indeed still the case. Second, even for moderately large values of $a_\mathrm{dd}$, reaching the saturation limit will require much lower particle numbers. This brings an observation of this characteristic property of quantum droplets within experimental reach. Third, due to the significantly lower densities, we expect that long-lived droplets could more easily be formed and studied using dipolar molecules, even if residual losses are present.

\section{Supersolid states}

We now move from single droplets in free space to supersolid arrays of multiple droplets in trapped samples. A supersolid is a counterintuitive state, where matter self-assembles into a crystal-like arrangement, but at the same time, still flows without friction. Such a state is thus characterized by a simultaneous breaking of gauge invariance and translation symmetry.  Following an ongoing, decade-long search for supersolids in solid helium ${}^4$He~\cite{Balibar2010,Boninsegni2012}, they have recently been observed in magnetically dipolar quantum gases~\cite{Bottcher2019,Tanzi2019,Chomaz2019,Bottcher2020}. 

In addition to the experimentally observed droplet supersolids, it has been shown theoretically that approaching the saturation density limit is connected to the emergence of other exotic supersolid states~\cite{Zhang2019,Hertkorn2021pattern,Zhang2021phases}. Rather than being formed by droplet arrays, these states are characterized by more complex density patterns. This behavior is closely related to pattern formation dynamics in other non-linear systems, ranging from classical ferrofluids and biological systems to geological structures and the physics of neutron stars~\cite{Cross1993}. 

However, as for the saturation of individual droplets, these more complex states are so far out of reach for experiments with magnetic atoms, due to limited atom numbers. As the decrease of the saturation density observed above in free space should approximately translate into trapped systems, we intuitively expect these states to be more accessible in molecular systems. 

In the following, we confirm this intuition, by studying the possible ground states of a molecular BEC in a harmonic trapping potential. We consider an oblate potential with trap frequencies $\omega_{x,y,z}=2\pi\times(100,100,200) \, \mathrm{Hz}$, and map out the phase diagrams for various dipolar lengths $a_\mathrm{dd}$. Scaling relations allow to easily generalize the results for this trap to other sets of parameters~\cite{Hertkorn2021pattern}. 

The results of our simulations are shown in Fig.~\ref{fig:phasediagram}. We find a wide variety of different phases, which include the well known droplet supersolids, but also honeycomb and pumpkin phases, as well as nearly degenerate ring and labyrinth states, which have recently gained significant interest~\cite{Zhang2019,Hertkorn2021pattern,Zhang2021phases}. Notably, we find that these more exotic phases become more dominant in the phase diagram as $a_\mathrm{dd}$ is increased.

More specifically, for sufficiently large scattering lengths $a_s$ the system is in the BEC phase for all $a_\mathrm{dd}$. Reducing $a_s$ leads to a softening of rotonic modes that triggers the breaking of the translational symmetry~\cite{Hertkorn2021SSD2D}. Still within the BEC regime this yields blood-cell-like ground states~\cite{Ronen2007}, where molecules accumulate on the outer rim of the BEC (Fig.~\ref{fig:phasediagram}b). For even lower $a_s$, crystalline patterns with length scales associated to the inverse roton momentum appear. Interestingly, the critical scattering length required to form these symmetry-broken states decreases with higher molecule number, which is due to the more dominant role played by quantum fluctuations at higher densities~\cite{Hertkorn2021pattern}. 

For small values of $a_\mathrm{dd}$ that are comparable to the situation in magnetic quantum gases, the phase diagram is largely dominated by supersolid and isolated arrays of droplets arranged in triangular patterns. An example density distribution is shown in Fig.~\ref{fig:phasediagram}c. At these small values of $a_\mathrm{dd}$, honeycomb patterns (Fig.~\ref{fig:phasediagram}d) emerge only at very high molecule numbers and scattering lengths. The emergence of these patterns can intuitively be connected to the density saturation of the droplets forming a droplet array. Once saturated, they can only respond to an increase in particle number by growing in size. For sufficiently high particle number, it becomes energetically favorable for the system to invert its density distribution, accumulate density in between the original droplet locations, and thus form a honeycomb pattern instead. 

In between these two extremes, the system can form a pattern that consists of elongated density stripes (Fig.~\ref{fig:phasediagram}e,f). Depending on the exact parameters, these stripes can be straight or curved, forming labyrinth structures, rings or pumpkins (Fig.~\ref{fig:phasediagram}g). As previously discussed, we find that the different morphologies of the labyrinth patterns are near degenerate in energy~\cite{Hertkorn2021pattern}, and only slightly lower in energy than the stripe patterns. A particular interesting question to address experimentally in this case is thus, whether the large scale degeneracies between different labyrinth patterns are robust, i.e. whether the labyrinths  really exist, or whether more symmetric stripe states with slightly higher energy dominate.

\begin{figure}[tb]
\centering
\includegraphics[width=0.44\textwidth]{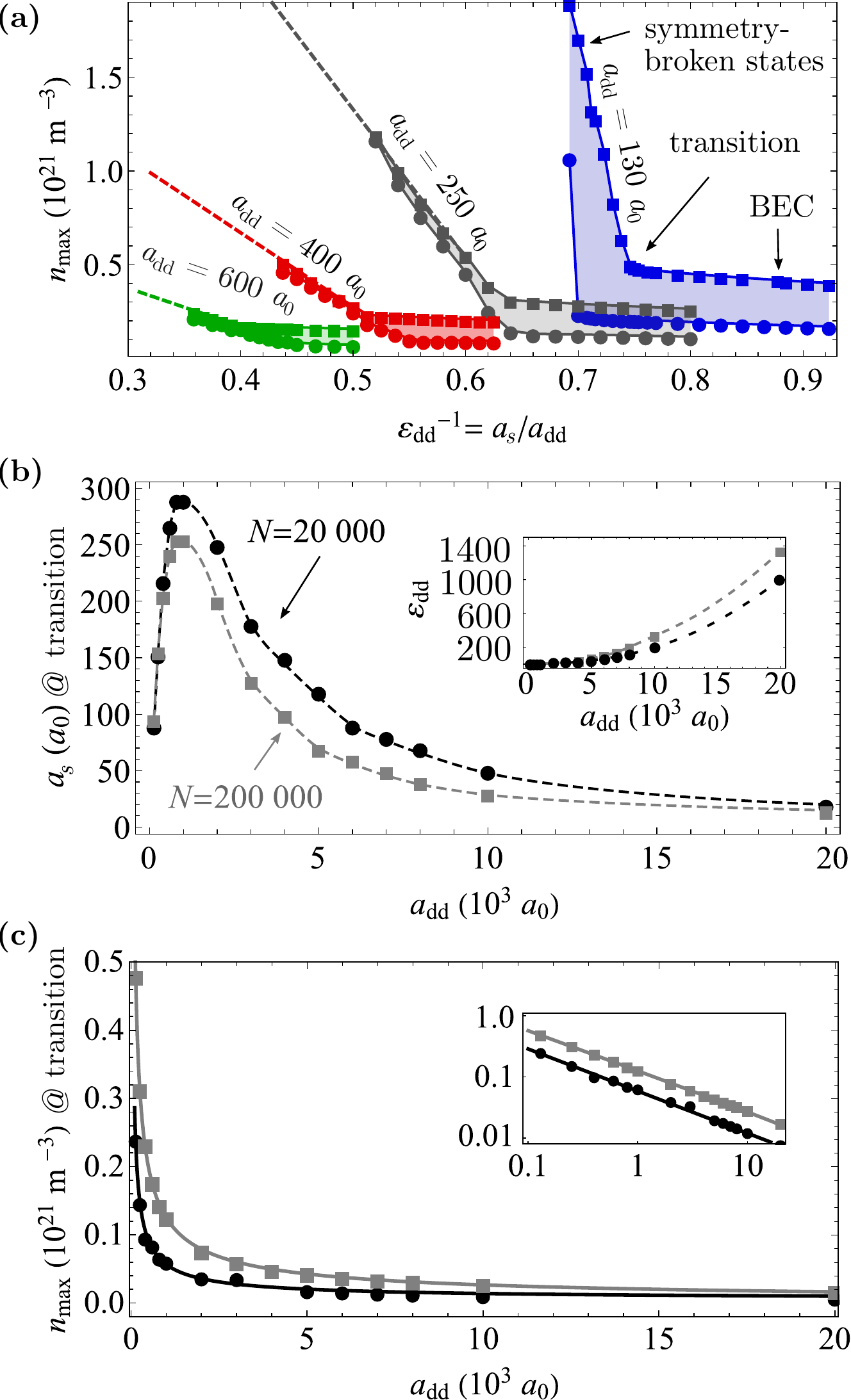}  
\caption{Transition from BEC to symmetry-broken states. (a) Peak density $n_\mathrm{max}$ of the states in the phase diagrams from Fig.~\ref{fig:phasediagram}. The circles (squares) denote states with $N=20\times 10^3$ ($200 \times 10^3$), respectively, and correspond to vertical cuts through the phase diagrams. The kink in the curves determines the transition between BEC and symmetry-broken states. While the density increases for larger molecule number in the BEC, the difference between states with different $N$ vanishes for the symmetry-broken states, indicating the presence of density saturation. (b) The scattering length required for the transition from a BEC to the symmetry-broken states shows a non-monotonous behavior with a maximum at $a_\mathrm{dd}\sim1000\,a_0$. The inset shows the corresponding required $\varepsilon_\mathrm{dd}$. (c) The peak density at the transition point decreases strongly with $a_\mathrm{dd}$ for both molecule numbers. The solid lines are power law fits, revealing a so far unknown scaling property of the eGPE. The inset is a double-log plot of the same data, further highlighting the power law behavior. All dashed lines are a guide to the eye. }
\label{fig:peakdensitytrap}
\end{figure}

As $a_\mathrm{dd}$ increases, the transitions to the various patterns shift towards significantly smaller $N$, from initially over several hundred thousand particles to a few thousand particles. This brings them within the realm of realistically achievable particle numbers. We anticipate that experiments mapping out the phase diagrams would, beyond being of fundamental interest, constitute another sensitive test of LHY beyond mean-field effects.

The systematic reduction of the peak density in the trapped system for larger $a_\mathrm{dd}$ is further highlighted in Fig.~\ref{fig:peakdensitytrap}, where we study it for two different example molecule numbers ($20\times$ and $200\times10^3$, respectively). In the BEC regime an increase in molecule number leads to an increase in peak density, as expected for a gas. Moreover, there is only a weak dependence of the peak density on $\varepsilon_\mathrm{dd}$. Both observations are well captured also by a variational approach for dipolar BECs (see Appendix). 

As the transition to the symmetry-broken states is crossed, the peak densities start to increase significantly and the difference between states with different $N$ quickly approaches zero, indicating ---again --- a liquid-like saturation of the density. We find that the values for $n_\mathrm{max}$ in this saturated regime are within a factor of $2$ of the saturation density $n_\mathrm{sat}$ predicted from the variational ansatz~\cite{Ferrier-Barbut2016}, which we previously discussed for free droplets in Fig~\ref{fig:dropletdensities}d. The residual difference can be attributed to the presence of the confinement and the different geometries of the various symmetry-broken states, which are not taken into account in this simple ansatz. Remarkably, despite the vastly different patterns formed by the individual states, there thus appears to exist one almost universal $n_\mathrm{sat}(a_s,a_\mathrm{dd},\omega_{x,y,z})$ (see Appendix).

The scattering length $a_s$ and dipolar parameter $\varepsilon_\mathrm{dd}$ required for the transition are plotted in Fig.~\ref{fig:peakdensitytrap}b as a function of $a_\mathrm{dd}$. For both molecule numbers studied, we observe first an increase of the required $a_s$, which approaches a maximum value at $a_\mathrm{dd}\sim1000\,a_0$, followed by a decrease for larger  $a_\mathrm{dd}$. This decrease can again be associated with more dominant quantum fluctuations. 

The peak density $n_\mathrm{max}$ at the transition point is shown in Fig.~\ref{fig:peakdensitytrap}c. We observe a continuous decrease with increasing $a_\mathrm{dd}$, which is described well by power law decays of the form $n_\mathrm{max} = A\, a_\mathrm{dd}^{-B}$ over the whole range of $a_\mathrm{dd}$ studied, with $A=(11.99\pm 0.18)$ and $B=(0.66\pm 0.01)$ for $N=200\times10^3$, and $A=(6.78\pm 0.72)$ and $B=(0.69\pm 0.02)$ for $N=20\times10^3$, respectively. This behavior indicates the presence of a previously unknown scaling property of the eGPE. Once the critical scattering lengths for the transition (see Fig.~\ref{fig:peakdensitytrap}b) are known, this density scaling can be reproduced by a variational approach for a dipolar BEC (see Appendix).

Extrapolating this decay to larger values beyond $a_\mathrm{dd}=2\times 10^4\,a_0$ one approaches a limit where the peak density decreases to values in the $10^{18}\,\mathrm{m}^{-3}$ range, three orders of magnitude lower than in magnetic quantum gases. In this case, the corresponding mean particle spacing becomes comparable to the characteristic length scale of the crystal structure, indicating that a situation with only one molecule per unit cell may be reached. The precise value of $a_\mathrm{dd}$ required to reach this limit depends on the details of the trapping potential, which sets the characteristic length scale of the symmetry-broken states~\cite{Hertkorn2021pattern}. 

This behavior is reminiscent of the transition from droplet to defect-induced supersolidity, which has so far only been identified and studied in the idealized theoretical model system of soft-core bosons~\cite{Cinti2014}. 
It will be interesting to study further whether a similar transition takes place here as well and whether the states at large $a_\mathrm{dd}$ show genuine supersolid behavior~\cite{Lu2015,Buechler2007,Astrakharchik2007,Macia2016,Cinti2017,Kora2019}. As the eGPE may not be a reliable description of the molecular BEC anymore in this strongly-correlated limit, and the allowed peak densities may even decrease to values that are below the critical density required for Bose-Einstein condensation, this question needs to be addressed using complementary systematic Monte-Carlo simulations. Conceptionally, such a defect-induced supersolid would be similar to solid ${}^4$He, where the search for supersolidity has been ongoing for decades~\cite{Balibar2010,Boninsegni2012}. However, in contrast to helium, single-particle-sensitive imaging and manipulation of ultracold molecular gases~\cite{Liu2018,Anderegg2019tweezer} could facilitate the study of the influence of defects, doping, dimensionality and disorder on the formation and dynamics of such a supersolid on the most fundamental level.

\section{Controlling the scattering properties of a molecular gas}

\begin{figure}[t]
\centering
\includegraphics[width=0.42\textwidth]{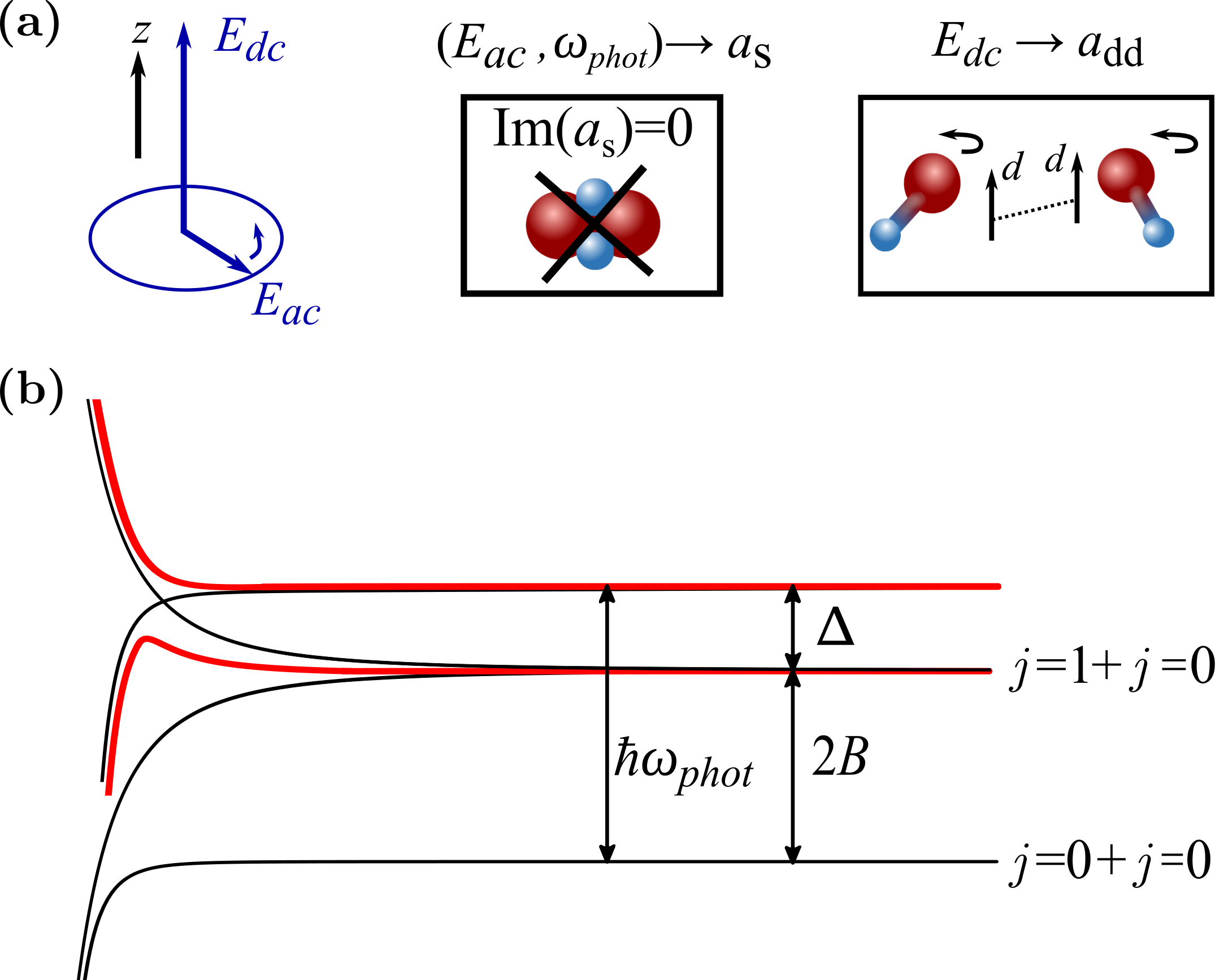} 
\caption{Collisional shielding and tuning of interactions. (a) The combination of a dc field $E_{dc}$, and a circularly polarized microwave ac field $E_{ac}$ with frequency $\omega_{phot}$ can be used to shield against collisional losses and independently control the dipolar length $a_\mathrm{dd}$ and the scattering length $a_\mathrm{s}$ between two ultracold molecules. The ac field and microwave frequency control $a_\mathrm{s}$, including the values $\operatorname{Re}(a_s)$ and $\operatorname{Im}(a_s)$, while the dc field controls $a_\mathrm{dd}$. (b) The microwave field is slightly blue detuned with respect to the  transition between the ground rotational state $j=0$ and the first excited state $j=1$ of a molecule. Here, $2B$ denotes the rotational level splitting and $\Delta = \hbar \omega_{phot} -2B$ is the detuning.  %
If the molecules in $j=0$ are dressed by the microwave, 
a repulsive barrier is created when they approach each other, as sketched by the red upper curve.
Access to the short-range region where molecules form a tetramer complex becomes forbidden,
suppressing losses.}
\label{fig:scheme}
\end{figure}

All of the physics investigated above strongly depends
on the capacity to fine-tune the dipolar length $a_\mathrm{dd}$
and the two-body s-wave scattering length $a_\mathrm{s}$
of an ultracold molecular gas.
In addition, in contrast to atoms, ground-state molecules can also lead 
to two-body collisional losses, no matter if the molecules are chemically reactive
or not~\cite{Croft2020,Quemener2021}.
As a result, the scattering length associated with molecular collisions
becomes a complex quantity, with an imaginary part 
directly linked to the magnitude of the molecular losses~\cite{Quemener2011}.
Therefore, tuning the scattering length of a gas of molecules 
implies to control {\it both} of its real and imaginary part.

\begin{figure}[tb]
\centering
\includegraphics[width=0.44\textwidth]{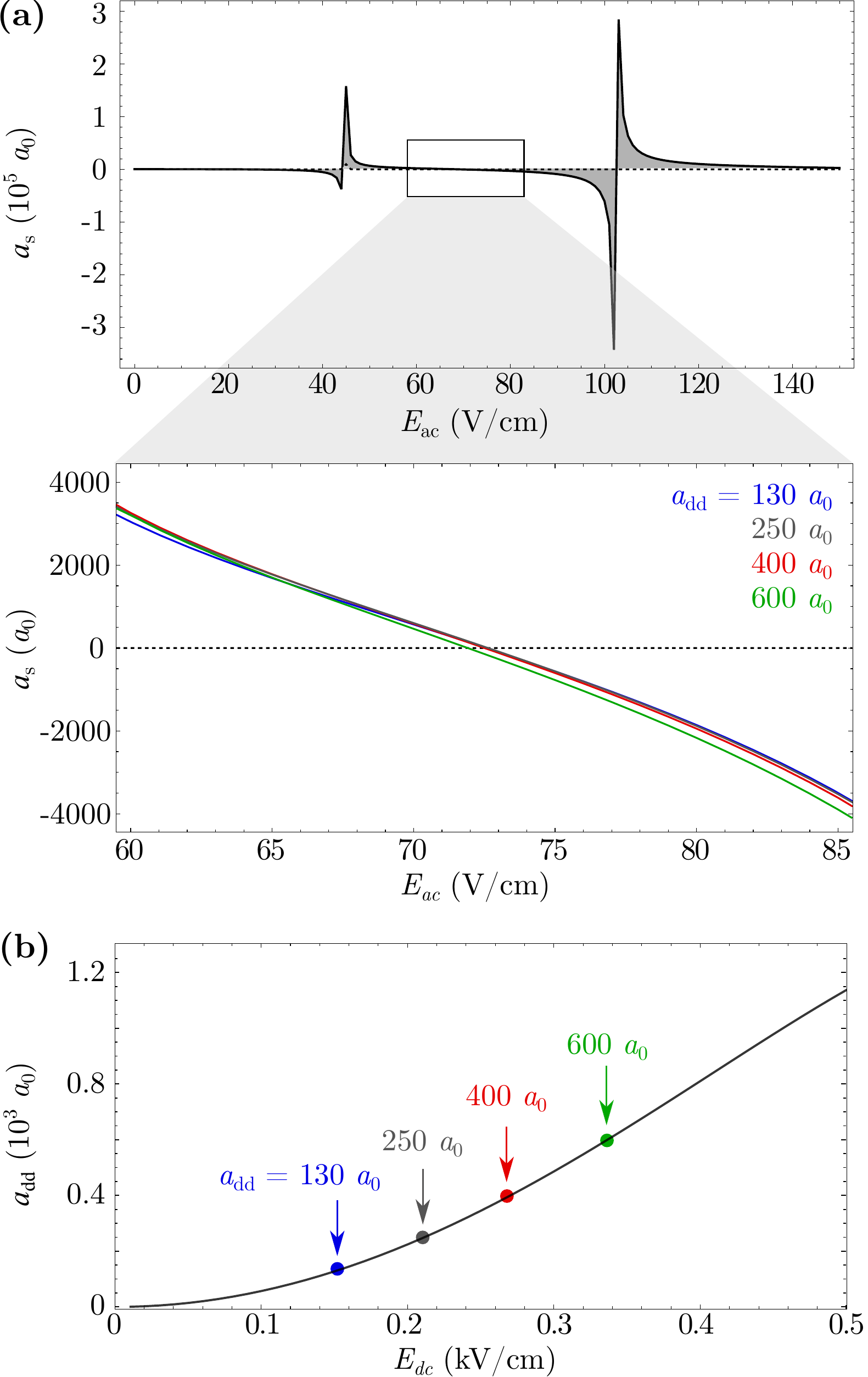} 
\caption{Independent tuning of contact and dipolar interactions via $E_{dc}$ and $E_{ac}$. (a) Tuning of $a_\mathrm{s}$ with $E_{ac}$
for $\operatorname{Re}(a_\text{s})$ (solid line)
and for $\operatorname{Im}(a_\text{s})$ (dotted line). Pronounced scattering resonances can be observed, which are the result of bound states that form in the long-range wells of molecules dressed by the microwaves. The inset provides a detailed view of the range $E_{ac} = [60 - 85]\,$V/cm, which covers all values of $\operatorname{Re}(a_\mathrm{s})$ used in this study. The corresponding values of $\operatorname{Im}(a_\text{s})$ in this field range are $\sim 0.04\,a_0$, meaning that molecular two-body losses are effectively suppressed on relevant timescales (see Appendix). The different solid lines show the effect of different values of the dc field $E_{dc}$ used in (b) on the scattering length $\operatorname{Re}(a_\text{s})$. This effect is  weak, demonstrating that the tuning parameters $E_{dc}$ and $E_{ac}$ can be considered as almost independent.
(b) Tuning of $a_\mathrm{dd}$ with $E_{dc}$. The value of $E_{ac}$ is set to $71$~V/cm.  The arrows indicate the dc fields needed to obtain the values of $a_\mathrm{dd}$ used in this study.  All highlighted parameters are summarized in Tab.~\ref{tab:asadd}.}
\label{fig:asadip}
\end{figure}

\subsection{Choice of shielding mechanism}

Molecular losses are an undesired outcome of a collision
as they will decrease the lifetime of the molecular gas.
This can be seen in Eq.~\ref{eq:eGPE}, where 
the eGPE becomes in general complex due to the complex scattering length,
which will provide the wavefunction an imaginary part 
associated with a decay process.
In a first step, it is then of crucial importance to prevent molecular losses
to occur for the physics described above to prevail.

For that purpose, different theoretical and experimental studies
have investigated the important question of
shielding molecules against losses.

For example, the use of a dc electric field in a confined geometry can protect
dipolar molecules in their ground rotational state $j=0$ from losses~\cite{Ticknor2010,Quemener2010,Micheli2010,Simoni}. The field polarizes the molecules such that they can, given the confined geometry, only interact through the repulsive part of the long-range dipolar interaction, which shields them from short-range losses. This was observed experimentally in a one-dimensional optical lattice 
for KRb molecules~\cite{DeMiranda}, and later also for magnetic Feshbach molecules of Er$_2$, where the electric field is replaced by a magnetic field~\cite{Frisch2015}.

However, this method is restricted to confined geometries only. In order to facilitate shielding also in more general geometries, such as the ones considered in Secs. III and IV, one can, counterintuitively, prepare molecules not in their $j=0$ ground  state
but in their $j=1$ first excited rotational state.  At a particular dc electric field, the molecules are protected from short-range losses via the creation of a long-range repulsive barrier in the entrance channel~\cite{Avdeenkov2006,Wang2015,Quemener2016,Gonzalez2017}.
This was successfully observed in recent experiments with ultracold
KRb molecules, increasing the lifetime of the gas to several seconds~\cite{Matsuda2020} and enabling 
evaporative cooling to take place~\cite{Li2021}.

Another example of efficient shielding in arbitrary geometries consists in using appropriately circular, blue-detuned microwaves~\cite{Gorkshkov2008,Lassabliere2018,Karman2018,Karman2019,Karman2020} on ground rotational state molecules. 
In addition to shielding, this method enables the desired control of both the real and imaginary parts of the scattering length~\cite{Lassabliere2018}. A recent experiment using CaF molecules in optical tweezers observed loss suppression with this method~\cite{Anderegg2021}. 

Finally, an optical shielding method was proposed~\cite{Xie2020} based on similar grounds, except that the microwave is replaced by an optical field and the mechanism relies on excited electronic rather than excited rotational states.

There are thus various established ways to shield molecular gases against losses and to tune their interactions. For the purpose of the study, we choose microwave shielding, as both parts of $a_\mathrm{s}$ can be tuned by two independent knobs, namely
the photon energy $\omega_{phot}$  and the amplitude of the ac field $E_{ac}$.
In addition, we will reserve a dc field to induce a dipole moment $d$ in the laboratory frame~\cite{Wang2015}. This creates a dipole-dipole interaction between the molecules characterized by $a_\mathrm{dd}$. 

The experimental setup is then a combination of these two fields, as sketched in Fig.~\ref{fig:scheme}.
Based on this setup, we will illustrate in the following how to tune the dipolar length and the scattering length in an ultracold molecular gas, taking rotational ground state NaRb molecules in $j=0$~\cite{Guo2016,Ye2018,Guo2018prx} as an example. 

\subsection{Suppression of molecular losses with $\omega_{phot}$}

As discussed above, the molecular losses are directly linked to 
$\operatorname{Im}(a_s)$.
To control this property we apply a microwave with circular polarization (Fig.~\ref{fig:scheme}b),
slightly blue-detuned between the ground rotational state $j=0$ 
and the first excited state $j=1$, with a detuning 
$\Delta = \hbar\omega_{phot} - 2B \sim  0.034 \, B$, $B/\hbar= 2\pi\times2.089\,$GHz being the rotational constant for NaRb~\cite{Guo2018rapid}. This value for the detuning is large enough to prevent collisions efficiently, but also sufficiently small not to influence the potential in which the molecules are trapped~\cite{Lassabliere2018}. The microwave frequency corresponds to $\omega_{phot} \sim 2\pi\,\times\,4.25$~GHz.
For these parameters, microwave dressing leads to a repulsive barrier that effectively suppresses losses for strong enough dressing. The suppression occurs typically for $E_{ac} \sim 40$~V/cm or larger, corresponding to Rabi frequencies  $\Omega = d \, E_{ac} / \hbar \sim 2\pi\,\times\,64$~MHz. This value is on the same order of magnitude as values that have already been reached in recent experiments~\cite{Anderegg2021}. 

The resulting $\operatorname{Im}(a_\text{s})$ is plotted in Fig.~\ref{fig:asadip}a as a function of $E_{ac}$. One can see that it remains close to zero, as expected for efficient microwave shielding.
We have checked that the residual value of $\operatorname{Im}(a_\text{s})\sim0.04\,a_0$ in the region of interest does not significantly affect the numerical results presented in the previous sections (see Appendix).

\begin{table}[t]
	\begin{tabular}{ccccc}
$d (D)$ & $a_\mathrm{dd}$ $(a_0)$& $a_\mathrm{s}$ $(a_0)$  & $E_\mathrm{dc}\,$(V/cm) & $E_\mathrm{ac}\,$(V/cm)\\ 
		\toprule
0.111&130 &  78 & 151.3  & 72.16 \\ 
\hline 
0.155&250 &  150 & 211.3  & 71.98 \\ 
\hline 
0.196&400 &   240 & 270.2 & 71.50 \\ 
\hline 
0.240& 600 &  360 & 336.6 & 70.40 \\ 
\hline
\end{tabular} 
\caption{Example values for the tuning of $a_\mathrm{dd}$ and $a_s$. The values correspond to those required to stabilize self-bound droplets with $2000$ molecules (see Fig.~\ref{fig:criticalatomnumber}). The microwave frequency used is $\omega_{phot} \sim 2\pi\,\times\,4.25$~GHz.}
\label{tab:asadd}
\end{table}

\subsection{Tuning the scattering length with $E_{ac}$}

Now that
$\operatorname{Im}(a_\text{s}) \to 0$,
we focus on the real part of the scattering length. 
Depending on the value of $E_{ac}$,
it was shown in~\cite{Lassabliere2018}
that 
$\operatorname{Re}(a_\text{s})$
can be controlled at will
to a desired value,
positive or negative, small or large. This is possible due to scattering resonances that arise from the formation of bound states in the long-range wells of two molecules~\cite{Avdeenkov2003,Avdeenkov2004} that are created and controlled by the microwave dressing amplitude $E_{ac}$. 

This tunability is illustrated in Fig.~\ref{fig:asadip}a. For the particular values of 
the real part of the scattering length considered in our study, we focus on the ac field range 
$E_{ac} = [60 - 85]$~V/cm, which corresponds to scattering length values between $-4000$ and $+4000$~$a_0$. 
The value of $E_{ac}$ then
sets directly the specific value of the scattering length. In Tab.~\ref{tab:asadd}, we summarize
example values of $E_{ac}$ required to cover the range of values of $a_s$ that were considered in the previous sections. 

\subsection{Tuning the dipolar length with $E_{dc}$}

The third knob is the dc field $E_{dc}$. It is used to induce an electric dipole moment $d$ in the laboratory frame, which sets the value of the dipolar length $a_\mathrm{dd}$. 

In previous work~\cite{Lassabliere2018}, microwave shielding was studied in the absence of  $E_{dc}$, such that there is no induced dipole moment and the dipolar length is zero. In this work we extend this approach to also allow for finite values of $E_{dc}$ using the theoretical formalism developed in Ref.~\cite{Wang2015}. 

The maximum value of $E_{dc}$ for which the shielding mechanism is active, and hence, the maximum achievable dipolar interaction strength, is set by the Stark shifts of the rotational states involved. Usually, in Fig.~\ref{fig:scheme} the $j=0+j=0$ channel has a stronger Stark shift than the  $j=1+j=0$ channel. Shielding using $E_{ac}$ is possible until these two channels cross each other. In our example for NaRb, this occurs for $E_{dc} \sim 0.5\,$kV/cm, corresponding to $a_\mathrm{dd}\sim 1200\,a_0$. The precise maximum value of $a_\mathrm{dd}$ that can be reached depends on the specific molecular species and the applied fields. It will thus be an interesting problem to identify appropriate combinations of $E_{dc}$, $E_{ac}$, and $\omega_{phot}$ that create --- at the same time --- arbitrary large values of $a_\mathrm{dd}$, arbitrary corresponding values of  $\operatorname{Re}(a_\text{s})$, while still maintaining efficient shielding with vanishing  $\operatorname{Im}(a_\text{s})$.

Fig.~\ref{fig:asadip}b shows the dipolar length as a function of $E_{dc}$, calculated for an example of value 
$E_{ac} =71$~V/cm. Other values of $E_{ac}$ provide similar curves. Conversely, the effect of the dc field on the scattering length
is highlighted in the inset of Fig.~\ref{fig:asadip}a,
where we plot $\operatorname{Re}(a_\text{s})$
as a function of $E_{ac}$
for different values of $E_{dc}$.
It can be seen that changing $E_{dc}$ does not affect
$\operatorname{Re}(a_\text{s})$ in a significant way,
such that one can consider
the tuning parameters $E_{dc}$ and $E_{ac}$ as almost independent.
In other words, in an experiment, one can first set the dc field to fix an appropriate
value of $a_\mathrm{dd}$, then one can set the ac field to fix
the appropriate value of $\operatorname{Re}(a_\text{s})$.

This enables a degree of tunability of the interactions that goes far beyond the possibilities in magnetic atoms. As summarized in Tab.~\ref{tab:asadd}, the values of $E_{dc}$ and $E_{ac}$ required to explore new regimes of the physics of dipolar BECs are well within the reach of current experiments. We are thus now at a turning point where — after decades of efforts — experiments with ultracold molecules are reaching a level of maturity, which brings the exploration of supersolidity and other dipolar quantum matter within reach.

\section{Conclusion and Outlook}
We have demonstrated that BECs of dipolar molecules allow to address a large variety of open questions regarding the conditions for the existence of dipolar droplets, supersolids and the properties of these states.  In comparison to magnetic atoms, this is due to the much larger and tunable electric dipole moments of the molecules. 

By combining dc and ac fields, we 
obtained the appropriate parameters to stabilize molecular Bose--Einstein condensates against losses and set the ideal conditions needed
for the emergence of self-bound droplets and supersolids. We anticipate these results to serve as a blueprint for near-future experiments and expect them to become an important benchmark when being systematically compared to the results of these experiments.  

In the same spirit of the two-body shielding mechanisms employed in this study, 
three-body shielding of molecules would be an additional remarkable feature compared to atoms. 
If three-body losses were also suppressed with similar mechanisms,
then exceptionally long-lived systems could be formed. First experiments on the static electric field shielding of ultracold KRb molecules have already suggested that this indeed appears to be the case~\cite{Matsuda2020,Li2021}. Theoretical investigations to further elucidate this question are currently in progress.

As classical ferrofluids are model systems for pattern formation in equilibrium, it will be interesting to explore whether strongly dipolar BECs can play a similar role in the quantum world. Moreover, our results in the strongly dipolar limit bring up questions regarding the role of \textit{discreteness} in the pattern formation dynamics. Is dipolar supersolidity only possible in the many-body limit of a droplet supersolid or is there a transition to defect induced-supersolidity? In this context it will also be interesting to explore similarities with the recently observed structured states in confined fermionic ${}^3$He~\cite{Shook2020} and ${}^4$He films on graphite substrates~\cite{Choi2021}, in order to establish a comprehensive understanding of supersolidity. 

\section*{Acknowledgments}
We are indebted to Tilman Pfau for generous support. We thank Hans Peter B\"uchler, Jens Hertkorn and the Stuttgart \textit{Dipolar Quantum Gases} team for fruitful discussions. We acknowledge Matthias Wenzel for developing the code used in our eGPE simulations, and Phillip Gro\ss\, for a critical reading of the manuscript. This project has received funding from the European Research Council (ERC) under the European Union’s Horizon 2020 research and innovation programme (Grant agreement No. 949431), the Vector Stiftung, the RiSC programme of the Ministry of Science, Research and Arts Baden-W\"urttemberg, and the Carl Zeiss Foundation. 
L. L. and G. Q. acknowledge funding from the FEW2MANY-SHIELD Project 
No. ANR-17-CE30-0015 from Agence Nationale de la Recherche.

\section*{Appendix}
\subsection{Two-body loss dynamics}
In order to reveal the influence of the residual $\operatorname{Im}(a_s)=0.04\,a_0$ we include an imaginary scattering length in the description of the mean-field interaction in Eq.~\ref{eq:eGPE} and study the resulting dynamics for a single, self-bound droplet state with $a_\mathrm{dd}= 130\,a_0$ and $\operatorname{Re}(a_s) = 100\,a_0$. Such a droplet has a peak density of around $0.5\,\times\,10^{21}\,\mathrm{m}^{-3}$ and thus constitutes a worst-case scenario, in which the high density leads to particularly fast losses. All other states considered, in particular for higher values of $a_\mathrm{dd}$, will be characterized by lower densities and thus exhibit slower, less significant losses. 

\begin{figure}[htb]
\centering
\vspace{15pt}
\includegraphics[width=0.49\textwidth]{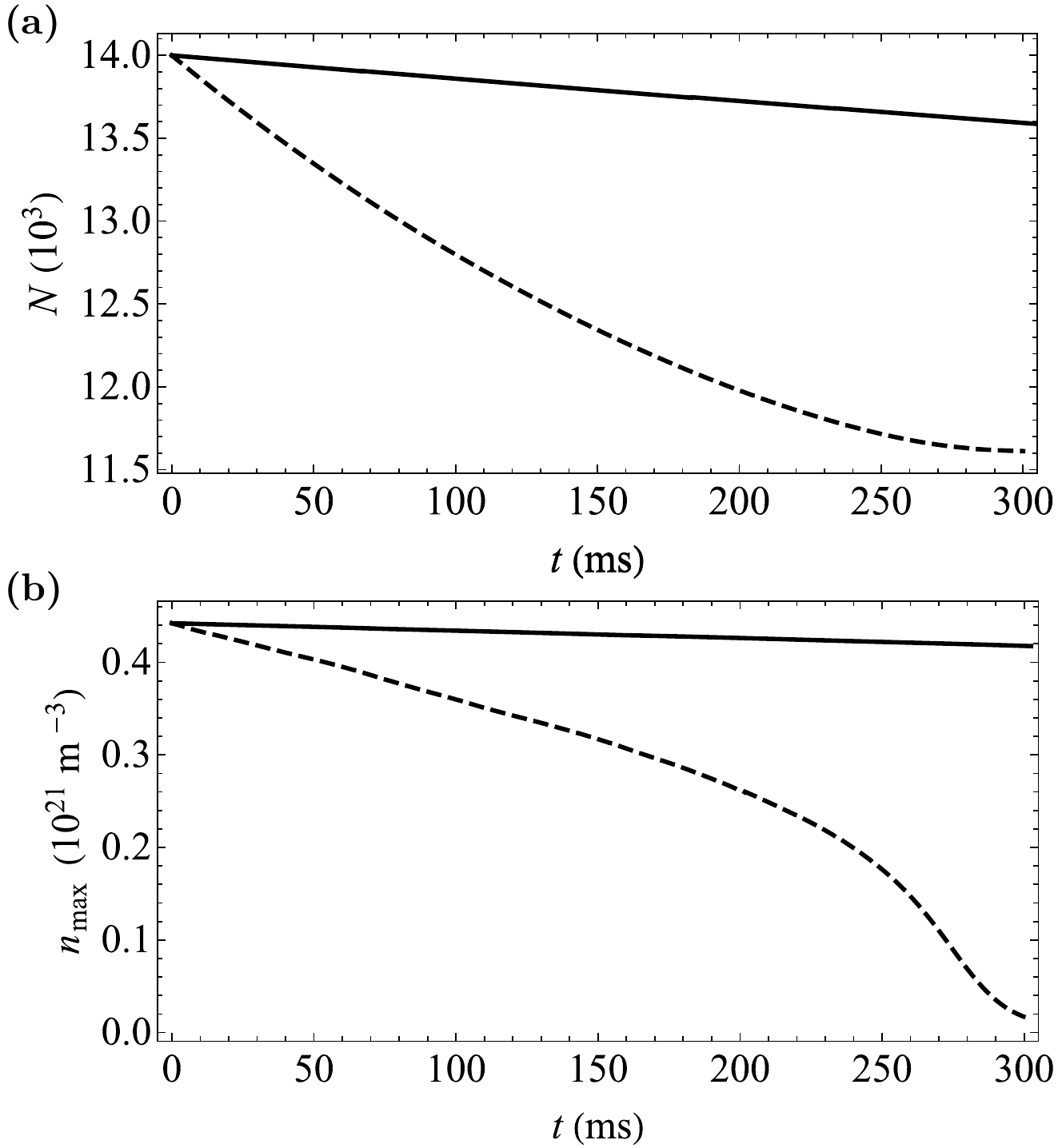} 
\caption{Loss dynamics in terms of (a) molecule number $N$ and (b) peak density $n_\mathrm{max}$ for a droplet state, with  $\operatorname{Re}(a_s) = 100\,a_0$ and $a_\mathrm{dd}= 130\,a_0$. The solid and dashed lines correspond to the expected value $\operatorname{Im}(a_s)=0.04\,a_0$ and a ten times larger value $\operatorname{Im}(a_s)=0.4\,a_0$, respectively.}
\label{fig:losses}
\end{figure}

The results of these simulations are shown in Fig.~\ref{fig:losses}. We observe a loss of less than $5\%$ of the molecules over $300\,$ms, consistent with efficient shielding. Even assuming less efficient shielding with a ten times higher value $\operatorname{Im}(a_s)=0.4\,a_0$, the timescale for the decay remains above $100\,$ms. It is thus still significantly longer than observed in similar droplets formed from dysprosium atoms, where the lifetime is limited by three-body losses~\cite{Bottcher2019droplet}. From this, we conclude that the residual two-body losses will play a minor role in the observation of the states discussed in the main text. 

\subsection{Density saturation in trapped systems}
\begin{figure}[htb]
\centering
\vspace{15pt}
\includegraphics[width=0.47\textwidth]{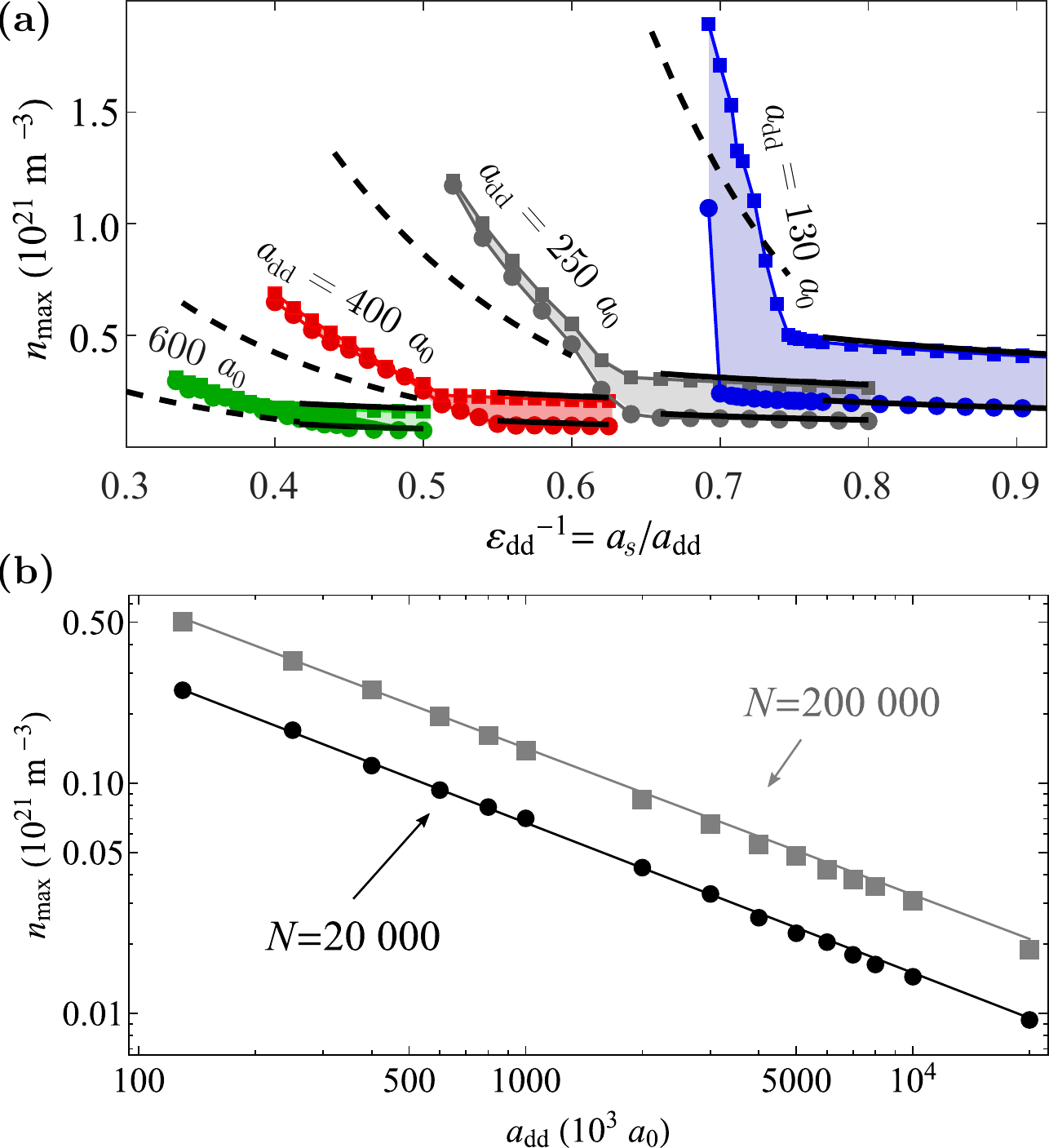} 
\caption{(a) Peak densities $n_\mathrm{max}$ across the transition from the BEC to the symmetry-broken states (see also Fig.~\ref{fig:peakdensitytrap}). The circles (squares) denote states with $N=20\times 10^3$ ($200 \times 10^3$), respectively. The solid black lines are obtained from a variational ansatz for dipolar BECs. The dashed black lines are derived from the variational result for free space droplets~\cite{Ferrier-Barbut2016}. For the latter, we have used $f_\mathrm{dip}=1$, with $f_\mathrm{dip}$ being the function appearing in the calculation of the dipolar mean-field
energy~\cite{Lahaye2009}. (b) The power law behavior of the density at the transition point, including the corresponding exponents, is well reproduced by the variational ansatz, if the correct values for $a_s$ and $a_\mathrm{dd}$, as obtained from eGPE simulations, are used as input parameters.}
\label{fig:densitiesAppendix}
\end{figure}

The density distribution of dipolar BECs can be well described by a variational ansatz using a Thomas-Fermi approximation~\cite{Lahaye2009}. For the range of the parameters $a_s$ and $a_\mathrm{dd}$ studied, the BEC is close to the transition to the symmetry broken states, and the LHY correction has to be included in the description~\cite{Lima2011}. Without doing so, no stable BEC solution exists in this parameter range. In Fig.~\ref{fig:densitiesAppendix}a we compare peak densities obtained via this procedure for $N=20\times10^3$ and  $N=200\times10^3$, respectively, to the results of our eGPE simulation, showing very good agreement. As discussed in the main text, there is a strong increase of the density with increasing molecule number, as expected for a gas. 

In order to recover the power laws observed in Fig.~\ref{fig:peakdensitytrap}c of the main text,  an evaluation of the peak densities exactly at the individual transition points is required. The variational ansatz is not capable of reproducing these points to high precision and analytical results are highly challenging, due to the anisotropic nature of the dipole-dipole interaction. However, using the interaction parameters obtained from our full eGPE simulation (see Fig.~\ref{fig:peakdensitytrap}b), the power laws can be recovered with high accuracy from the variational ansatz (Fig.~\ref{fig:densitiesAppendix}b). Moreover, given a single known solution at the transition point (obtained either from the eGPE or through measurements in an experiment) other solutions can again be found through the known scaling properties of the eGPE~\cite{Hertkorn2021SSD2D,Hertkorn2021pattern}. 

Finally, we compare the saturation density of a single droplet in free space with the peak densities observed for the various symmetry-broken states, as discussed in Fig.~\ref{fig:peakdensitytrap}a of the main text. We find that a simple variational ansatz for the droplets~\cite{Ferrier-Barbut2016} yields densities that are remarkably close to the ones observed for the much more complex symmetry-broken states. We interpret this as an indication for the existence of a general saturation density for dipolar quantum liquids. 

\newpage

\bibliography{biblio,refs,Citation_Paris}

\end{document}